\documentclass{jetpl}

\usepackage{graphicx}

\twocolumn
\lat

\DeclareMathOperator{\sgn}{sgn}

\title{Critical temperature of superconductor/ferromagnet bilayers}
\rtitle{Critical temperature of SF bilayers}
\sodtitle{Critical temperature of superconductor/ferromagnet bilayers}

\author{Ya.\,V.~Fominov$^{+*}$\/\thanks{e-mail: fominov@landau.ac.ru}, N.\,M.~Chtchelkatchev$^+$,
and A.\,A.~Golubov$^*$}
\rauthor{Ya.\,V.~Fominov, N.\,M.~Chtchelkatchev, and A.\,A.~Golubov}
\sodauthor{Fominov, Chtchelkatchev, Golubov}

\address{$^+$L.\,D.~Landau Institute for Theoretical Physics RAS, 117940 Moscow, Russia\\~\\
$^*$Department of Applied Physics, University of Twente, P.O.~Box~217, 7500 AE Enschede,
The~Netherlands}

\dates{8 June 2001}{*}

\abstract{Superconductor/ferromagnet bilayers are known to exhibit nontrivial dependence of the
critical temperature $T_c$ on the thickness $d_f$ of the ferromagnetic layer. We develop a
general method for investigation of $T_c$ as a function of the bilayer's parameters. It is shown
that interference of quasiparticles makes $T_c(d_f)$ a nonmonotonic function. The results are in
good agreement with experiment. Our method also applies to multilayered structures.}


\begin{document}
\maketitle

Recently, much attention has been paid to properties of hybrid proximity systems containing
superconductors (S) and ferromagnets (F); new physical phenomena were predicted and observed in
these systems \cite{Ryazanov,Kontos,Radovic,Tagirov_PRL}. One of the most striking effects in SF
layered structures is highly nonmonotonic dependence of the critical temperature $T_c$ of the
system on the thickness $d_f$ of the ferromagnetic layers. Experiments exploring this
nonmonotonic behavior have been performed previously on SF multilayers such as Nb/Gd
\cite{Jiang}, Nb/Fe \cite{Muhge}, V/V-Fe \cite{Aarts}, and Pb/Fe \cite{Lazar} but the results
(and, in particular, comparison between the experiments and theories) were not conclusive.

To perform reliable experimental measurements of $T_c(d_f)$, it is essential to have $d_f$ large
compared to the interatomic distance; this situation can be achieved only in the limit of weak
ferromagnets. Active experimental investigations of SF bilayers and multilayers based on Cu-Ni
dilute ferromagnetic alloys are carried out by the group of Ryazanov\footnote{Ryazanov, Oboznov,
Prokof'ev {\it et al.}~--- hereafter referenced as ROP.} \cite{Ryazanov_new}. In SF bilayers,
they observed highly nonmonotonic dependence $T_c(d_f)$. While the reason for this effect in
multilayers can be the $0$--$\pi$ transition \cite{Radovic}, in a bilayer system with a single
superconductor this mechanism is irrelevant and the cause of the effect is quasiparticle
interference specific to SF structures.

In the present paper, motivated by the ROP experiment \cite{Ryazanov_new} we study theoretically
the critical temperature of SF bilayers. Previous theoretical investigations of $T_c$ in SF
structures were concentrated on systems with thin or thick S(F) layers [compared to the coherence
length of the superconductor (ferromagnet)]; with SF boundaries having very low or very high
transparency; besides, the exchange energy was often assumed to be much larger then the critical
temperature \cite{Radovic,Aarts,Lazar,Buzdin,Demler,Khusainov,Tagirov}. The parameters of the ROP
experiment do not correspond to any of these limiting cases. In the present paper we develop an
approach giving opportunity to investigate not only the limiting cases of parameters, but also
the intermediate region. Using our method, we find different types of nonmonotonic behavior of
$T_c$ as a function of $d_f$ such as minimum of $T_c$ and even reentrant superconductivity
\cite{future}. Comparison of our theoretical predictions with the experimental data shows good
agreement.

We assume that dirty-limit conditions are fulfilled, and calculate the critical temperature of
the bilayer within the framework of the linearized Usadel equations for S and F layers (the
domain $0<x<d_s$ is occupied by the S metal, $-d_f<x<0$~--- by the F metal, see Fig.1). Near
$T_c$ the normal Green function is $G=\sgn\omega_n$, and the Usadel equations for the anomalous
function $F$ take the form:
\begin{gather}
\label{U_1}
\xi_s^2\, \pi T_{cs} \frac{d^2F_s}{dx^2}- |\omega_n| F_s +\Delta=0,\quad 0<x<d_s;\\
\xi_f^2\, \pi T_{cs} \frac{d^2F_f}{dx^2} -(|\omega_n|+iE_\mathrm{ex} \sgn\omega_n) F_f=0, \\
-d_f<x<0; \nonumber\\
\label{U_3}
\Delta\ln\frac{T_{cs}}{T} = \pi T \sum_{\omega_n} \left( \frac\Delta{|\omega_n|} -F_s \right),
\end{gather}
where $\xi_s=\sqrt{D_s/2\pi T_{cs}}$, $\xi_f=\sqrt{D_f/2\pi T_{cs}}$, $\omega_n=\pi T(2n+1)$ with
$n=0,\pm 1, \pm 2,\ldots$ are the Matsubara frequencies, $E_\mathrm{ex}$ is the exchange energy,
and $T_{cs}$ is the critical temperature of the S layer. $F_{s(f)}$ denotes the function $F$ in
the S(F) region.

Equations \eqref{U_1}-\eqref{U_3} must be supplemented with the boundary conditions at the outer
surfaces of the bilayer:
\begin{equation}
\frac{dF_s(d_s)}{dx}=\frac{dF_f(-d_f)}{dx}=0,
\end{equation}
as well as at the SF boundary:
\begin{align}
& \xi_s\frac{dF_s(0)}{dx} =\gamma\xi_f \frac{dF_f(0)}{dx}, &&
\gamma =\frac{\rho_s\xi_s}{\rho_f\xi_f}, \\
& \xi_f\gamma_b \frac{dF_f(0)}{dx} = F_s(0)-F_f(0), && \gamma_b
=\frac{R_b {\cal A}}{\rho_f\xi_f}. \label{bound_2}
\end{align}
Here $\rho_{s,f}$ are the normal state resistivities of the S and F metals, $R_b$ is the total
resistance of the SF boundary, and ${\cal A}$ is its area. The Usadel equation in the F layer is
readily solved:
\begin{gather}
F_f=C(\omega_n) \cosh\left( k_f (x+d_f)\right),\\
\text{with } k_f=\frac 1{\xi_f} \sqrt{\frac{|\omega_n|+iE_\mathrm{ex} \sgn\omega_n}{\pi T_{cs}}},
\end{gather}
and the boundary condition at $x=0$ can be written in closed form with respect to $F_s$:
\begin{equation}
\label{bound_Phi}
\xi_s \frac{dF_s(0)}{dx} =\frac\gamma{\gamma_b +B_f(\omega_n)}F_s(0),
\end{equation}
where $B_f(\omega_n)=\left[ k_f \xi_f \tanh (k_f d_f) \right]^{-1}$.

This boundary condition is complex. In order to rewrite it in real form, we do the usual trick
and go over to the functions $F^\pm =F(\omega_n)\pm F(-\omega_n)$. The symmetric properties of
$F^+$ and $F^-$ are trivial, so we will treat only positive $\omega_n$. The self-consistency
equation is expressed only via the symmetric function $F_s^+$:
\begin{equation}
\label{self_cons}
\Delta\ln\frac{T_{cs}}T = \pi T \sum_{\omega_n>0} \left(\frac{2\Delta}{\omega_n}-F_s^+ \right),
\end{equation}
and the problem of determining $T_c$ can be formulated in closed form with respect to $F_s^+$.
This is done as follows. The Usadel equation for $F_s^-$ does not contain $\Delta$, hence it can
be solved analytically. After that we exclude $F_s^-$ from the boundary condition
\eqref{bound_Phi} and arrive at the effective boundary conditions for $F_s^+$:
\begin{equation}
\label{bound_p}
\xi_s \frac{dF_s^+ (0)}{dx}= W(\omega_n) F_s^+ (0),\qquad\frac{dF_s^+ (d_s)}{dx}=0,
\end{equation}
where
\begin{gather}
\label{W_def}
W=\gamma \frac{A_s (\gamma_b+\Real B_f)+ \gamma}{A_s |\gamma_b+B_f|^2 +\gamma
(\gamma_b+\Real B_f)},\\
\nonumber
A_s= k_s \xi_s \tanh (k_s d_s),\quad\text{with } k_s=\frac 1{\xi_s}
\sqrt{\frac{\omega_n}{\pi T_{cs}}}.
\end{gather}
The self-consistency equation \eqref{self_cons}, the boundary conditions
\eqref{bound_p}-\eqref{W_def} together with the Usadel equation for $F_s^+$:
\begin{gather}
\label{usadel_b}
\xi_s^2\, \pi T_{cs} \frac{d^2 F_s^+}{dx^2}- \omega_n F_s^+ +2\Delta=0
\end{gather}
will be used below to find the critical temperature of the bilayer.

The Green function (in mathematical sense) of the problem \eqref{bound_p}--\eqref{usadel_b} can
be expressed via solutions $v_1$, $v_2$ of Eq. \eqref{usadel_b} without $\Delta$, satisfying the
boundary conditions at $x=0$ and $x=d_s$, respectively:
\begin{multline}
G(x,y;\omega_n)= \frac{k_s \xi_s / \omega_n}{\sinh( k_s d_s) +a\cdot\cosh\left( k_s d_s\right)}
\times\\
\times\left\{ \begin{aligned}
v_1(x) v_2(y), && x\le y\\ v_2(x) v_1(y), && y\le x
\end{aligned} \right. , \label{G}
\end{multline}
where $a=W(\omega_n)/ k_s \xi_s$ and
\begin{align}
v_1(x) & =\cosh( k_s x) +a\cdot\sinh( k_s x),\notag\\
v_2(x) & =\cosh\left( k_s (x-d_s) \right).
\end{align}
Having found $G(x,y;\omega_n)$, we can write the solution of
Eqs.~\eqref{bound_p}--\eqref{usadel_b} as
\begin{equation}
F_s^+ (x;\omega_n)=2\int_0^{d_s} G(x,y;\omega_n) \Delta(y) dy.
\end{equation}
Substituting this into the self-consistency equation \eqref{self_cons}, we obtain
\begin{multline}
\Delta(x) \ln\frac{T_{cs}}{T_c} =\\
= 2\pi T_c \sum_{\omega_n>0} \left[ \frac{\Delta(x)}{\omega_n}-
\int_0^{d_s} G(x,y;\omega_n) \Delta(y) dy \right]. \label{sc_Green}
\end{multline}
This equation can be expressed in symbolic form: $\Delta\ln(T_{cs}/T_c)=\hat L\Delta$. Then $T_c$
is determined from the condition
\begin{gather}
\label{det}
\det\left(\hat L- \hat 1 \ln\frac{T_{cs}}{T_c} \right)=0
\end{gather}
that Eq.~\eqref{sc_Green} has a nontrivial solution with respect to $\Delta$. Numerically, we put
our problem \eqref{sc_Green}--\eqref{det} on the spatial grid so that the linear operator $\hat
L$ becomes a finite matrix.

Equations \eqref{G}--\eqref{det} are our central result; substituting the concrete parameters of
the system we can easily find the critical temperature numerically and in certain cases
analytically. (The models considered previously
\cite{Radovic,Aarts,Lazar,Buzdin,Demler,Khusainov,Tagirov} correspond to the limiting cases of
our theory.)

We apply our method to fit the ROP experimental data \cite{Ryazanov_new}; the result is presented
in Fig.2. Estimating the parameters $d_s=11$\,nm, $T_{cs}=7$\,K, $\rho_s=7.5$\,$\mu\Omega$\,cm,
$\xi_s=8.9$\,nm, $\rho_f=60$\,$\mu\Omega$\,cm, $\xi_f=7.6$\,nm, $\gamma=0.15$ from the experiment
and fitting only $E_\mathrm{ex}$ and $\gamma_b$, we find good agreement between our theoretical
predictions and the experimental data. The fitting procedure was the following: first, we
determine $E_\mathrm{ex}\approx 130$\,K from the position of the minimum of $T_c(d_f)$; second,
we find $\gamma_b\approx 0.3$ from fitting the vertical position of the curve. The deviation of
our curve from the experimental points is small; it is most pronounced in the region of small
$d_f$ corresponding to the initial decrease of $T_c$. This is not unexpected because when $d_f$
is of the order of a few nanometers, the thickness of the F film may vary significantly along the
film (which is not taken into account in our theory) and the thinnest films can even be formed by
an array of islands rather then by continuous material. At the same time, we note that the
minimum of $T_c$ takes place at $d_f\approx 5$\,nm, when with good accuracy the F layer has
uniform thickness.

The position of the minimum of $T_c(d_f)$ can be estimated from qualitative arguments based on
interference of quasiparticles in the ferromagnet. Let us consider a point $x$ inside the F
layer. According to Feynman's interpretation of quantum mechanics \cite{Feynman}, the
quasiparticle wave function [we are interested in anomalous wave function of correlated
quasiparticles, which characterizes the superconductivity; this function is equivalent to the
anomalous Green function $F(x)$] may be represented as a sum of the wave amplitudes over all
classical trajectories; the wave amplitude for a given trajectory equals $\exp(iS)$, where $S$ is
the classical action along this trajectory. To obtain our anomalous wave function we must sum
over trajectories that (i) start and end at the point $x$, (ii) change the type of the
quasiparticle (i.e., convert an electron into a hole or vice versa). There are four kinds of
trajectories which should be taken into account (see Fig.1). Two of them (denoted 1 and 2) start
in the direction toward the SF interface (as an electron and as a hole), experience the Andreev
reflection and return to the point $x$. The other two trajectories (denoted 3 and 4) start in the
direction away from the interface, experience normal reflection at the outer surface of the F
layer, move toward the SF interface, experience the Andreev reflection there, and finally return
to the point $x$. The main contribution is given by the trajectories normal to the interface. The
corresponding actions are $S_1=-S_2=-Qx$ and $S_3=-S_4=-Q(2d_f+x)$ (note that $x<0$), where $Q$
is the difference between the wave numbers of the electron and the hole. To make our arguments
more clear, we assume that the ferromagnet is strong, the SF interface is ideal, and consider the
clean limit first: in this case $Q=k_e-k_h=\sqrt{2m(E+E_\mathrm{ex}+\mu)}-
\sqrt{2m(-E-E_\mathrm{ex}+\mu)}\approx 2E_\mathrm{ex}/v$, where $E$ is the quasiparticle
energy, $\mu$ is the Fermi energy, and $v$ is the Fermi velocity. Thus the anomalous wave
function of the quasiparticles is $F(x)\propto \sum_{n=1}^4 \exp(iS_n)\propto \cos(Qd_f)
\cos(Q(d_f+x))$. The suppression of $T_c$ by the ferromagnet is determined by the value of the
wave function at the SF interface: $F(0)\propto \cos^2 (Qd_f)$. The minimum of $T_c$ corresponds
to the minimal value of $F(0)$ which is achieved at $d_f=\pi/2Q$. In the dirty limit the above
expression for $Q$ is replaced by $Q = \sqrt{E_\mathrm{ex}/D_f}$, hence the minimum of $T_c(d_f)$
takes place at
\begin{equation}
d_f^{\rm (min)} = \frac\pi 2 \sqrt{\frac{D_f}{E_\mathrm{ex}}}.
\end{equation}
In the case of the ROP bilayer \cite{Ryazanov_new} we obtain $d_f^{\rm (min)}\approx 7$\,nm, whereas
the experimental value is $5$\,nm (Fig.2); thus our qualitative estimate appears to be
reasonable.

The method developed in this paper applies directly to multilayered SF structures (in particular,
to trilayers) in the $0$-state, where an SF bilayer can be considered as an elementary cell of
the system. A generalization can be made, which allows to take account of possible
superconductive and/or magnetic $\pi$-states.

In conclusion, we have developed a method for calculating the critical temperature of a SF
bilayer as a function of parameters of the junction. The approach developed here gives an
opportunity to evaluate $T_c$ in wide range of parameters. We demonstrate that there is good
agreement between the experimental data and our theoretical predictions. Qualitative arguments
are given, which explain the nonmonotonic behavior of the function $T_c(d_f)$. Extensive details
of our study will be published elsewhere \cite{future}.

We thank V.\,V.~Ryazanov and M.\,V.~Feigel'man for stimulating discussions and useful comments on
the manuscript. We are especially indebted to V.\,V.~Ryazanov for communicating the experimental
result of his group to us prior to the detailed publication. Also we are grateful to
M.\,Yu.~Kupriyanov and Yu.~Oreg for enlightening comments. Ya.V.F. acknowledges financial support
from the Russian Foundation for Basic Research (RFBR), project No. 01-02-17759, and from
Forschungszentrum J\"ulich (Landau Scholarship). The research of N.M.C. was supported by the
RFBR, project No. 01-02-06230, by Forschungszentrum J\"ulich (Landau Scholarship), by the
Netherlands Organization for Scientific Research (NWO), and by the Swiss National Foundation.

\vfill\eject

\begin{figure}
\centerline{\includegraphics[width=90mm]{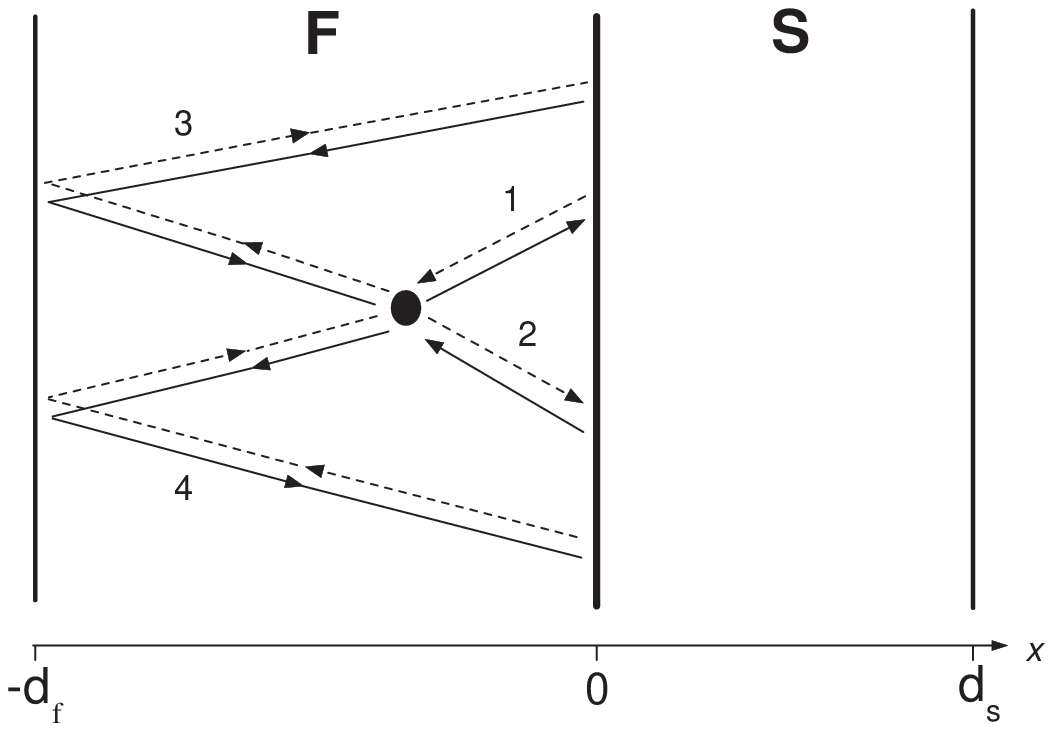}}
\caption{}
\end{figure}

\begin{figure}
\centerline{\includegraphics[width=90mm]{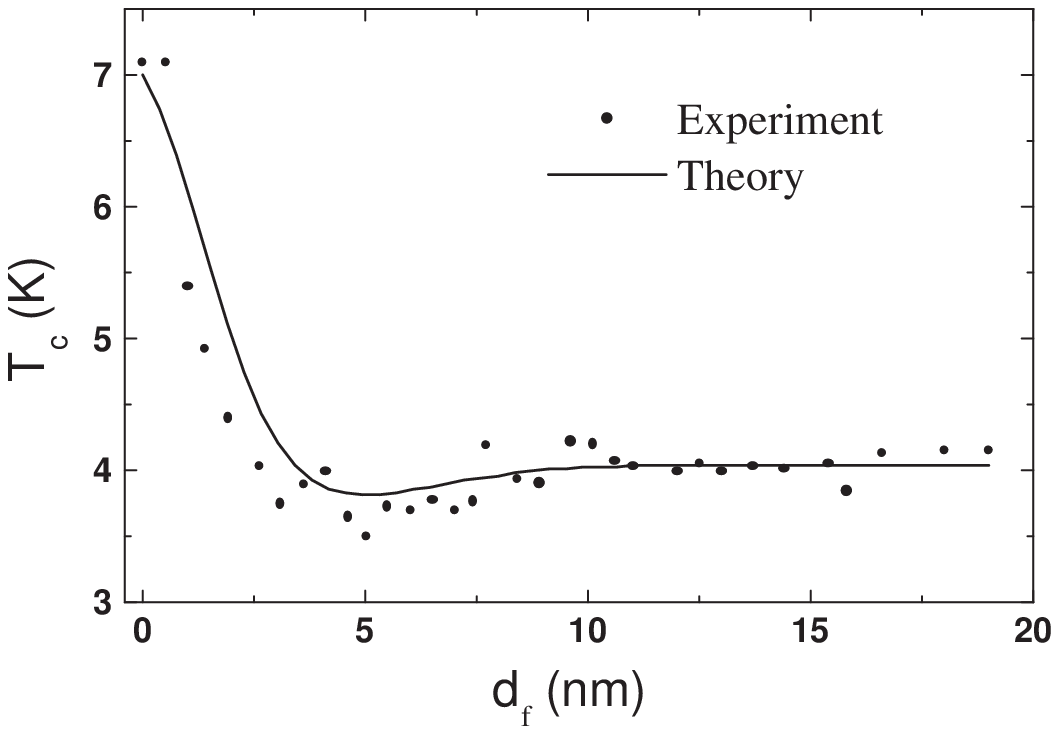}}
\caption{}
\end{figure}

Fig.1. The SF bilayer. The F and S layers occupy the regions $-d_f<x<0$ and $0<x<d_s$,
respectively. The four types of trajectories contributing (in Feynman path integral sense) to the
anomalous wave function of correlated quasiparticles are shown in the ferromagnetic region. The
solid lines correspond to electrons, the dashed lines~--- to holes; the arrows indicate the
direction of the velocity.

Fig.2. Theoretical fit to ROP's experimental data \cite{Ryazanov_new}. In the experiment, Nb was
the superconductor (with $d_s=11$\,nm, $T_{cs}=7$\,K) and Cu$_{0.43}$Ni$_{0.57}$ was the weak
ferromagnet. From our fit we estimate $E_\mathrm{ex}\approx 130$\,K and $\gamma_b\approx 0.3$.

\end{document}